# Pretransitional behavior of electrooptic Kerr effect in liquid thymol

Aleksander Szpakiewicz-Szatan[1,2,a,*], Sylwester J. Rzoska[1,b], Aleksandra Drozd-Rzoska[1,c]

[1]Institute of High Pressure Physics, Polish Academy of Sciences, ul. Sokołowska 29/37, 01-142 Warsaw, Poland

[2]Faculty of Physics, Warsaw University of Technology, ul. Koszykowa 75, 00-662 Warsaw, Poland

[a]orcid.org/0000-0002-4614-139X

[b]orcid.org/0000-0002-2736-2891

[c]orcid.org/0000-0001-8510-2388

[*]e-mail: aszpasza@mail.unipress.waw.pl

**Abstract**

Melting/freezing are canonical examples of discontinuous phase transitions, for which no pretransitional effects in the liquid phase are expected. For the solid phase, weak premelting effects are evidenced. This report shows long-range, critical-like, pretransitional effects in liquid thymol detected in electrooptic Kerr effect (EKE) studies. Notably is the negative sign of EKE pretransitional anomaly. Studies are supplemented by the high-resolution dielectric constant temperature-related scan, which revealed a weak premelting effect in the solid phase. Both EKE and dielectric constant show a 'crossover' change in the liquid phase, ca, 10 K above the freezing temperature. It can be recognized as the hallmark of the challenging Liquid-Liquid transition phenomenon.

## 1. Introduction

Melting/freezing discontinuous phase transition is one of the most common phenomena in nature, experienced in our everyday life [1]. It is also used in many technological processes, from food preservation [2] to physicochemical technologies [3–5]. Despite its significance, the cognitive insight into the phenomenon remains limited. It can be explained by inherent phenomenal limitations associated with melting/freezing, particularly the practical lack of pretransitional effects. Only weak and range-limited premelting changes for some physical properties are observed in the solid/crystalline side of the melting temperature $T_m$. This phenomenon is recognized as a particularly significant experimental reference and constitutes the base for so-called 'grain models' for melting/freezing, which dominated nowadays. They explain melting via the appearance of nano/micro- grains, separated by 'channels' with nano/micro constrained liquid [6, 7]. The latter's growth on heating destabilizes the crystalline solid and melts into the liquid state. Generally, no pretransitional effects are expected for the liquid side of $T_m$ [8, 9]. Notable that for continuous phase transitions, long-range pretransitional effects, also in the high-temperature ('liquid') domain, are the basic inspiration and refrence for the *Physics of Critical Phenomena*, one of the grand universalistic successes of 20[th] Century Physics [8, 9]. For melting/freezing discontinuous phase transition, such a cognitively beneficial phenomenon seemed absent [10, 11].

However, recently two striking exceptions from the rule 'no pretransitional effects in the liquid state' for melting/freezing have been reported [12-14]. First, the long-range pretransitional effect was registered in the liquid phase of cyclooctanol, using the electrooptic Kerr effect (EKE) and nonlinear dielectric effect (NDE) methods [12, 13]. Cyclooctanol is the plastic crystal-forming (PC) material associated with the solidification in the Orientationaly Disordered Crystal (ODIC) phase. However, in ref. [13] the model coherently explaining the mentioned pretransitional anomalies in ODIC-formers and in the isotropic liquid phase of rod-like liquid crystaline (LC) materials was proposed. The latter is known for decades and included in the canon of the *Critical Phenomena Physics* [10, 15-20]. Hence, melting/freezing in ODIC-forming (PC) materials should leave the family of materials with the canonic melting/freezing phenomenon. Consequently, the paradigm for the lack of the long-range pretransitional effects near the melting and freezing temperatures seemed to remain.

However, this year explicit strong and long-range EKE and NDE pretransitional effects in liquid menthol have been reported [14]. In menthol, the 'classic' melting/freezing discontinuous transition occurs. There are no mesophases (LC or PC) between liquid and crystalline phases.

This report presents results of electrooptic Kerr effect studies, supplemented by dielectric constant scan, in thymol, i.e., the material from the same homologous series of terpenoids as menthol. Studies revealed remarkable pretransitional changes but significantly different from the ones observed in menthol [14]. Worth recalling is the practical significance of thymol. It is a natural (i.e., extracted from plants) compound widely used in cosmetic and pharmaceutical industries [21, 22].

## 2. Experimental

Results presented in this report are for thymol, also known as 2-isopropyl-5-methylphenol ($C_{10}H_{14}O$), the natural monoterpenoid phenol. The tested sample of solid thymol (Biosynth, min 98% purity) was heated just above the melting point of 323 K (as stated by producer's datasheet, endothermic peak between 313 and 333 K in DSC measured by Trivedi et al. [23] and our dielectric measurements). It was tested via broadband dielectric spectroscopy (BDS) using Novocontrol impedance analyzer, supported by Quattro temperature unit and subsequently using the electrooptic Kerr effect (EKE birefringence: B). The target of BDS studies was to obtain high-resolution dielectric constant data in a broad range of temperatures to supplement EKE-related results. Figure 1 shows spectra for the real part of dielectric permittivity in thymol, showing that frequency $f = 10 kHz$ is located in the mid of the static domain, which justifies the assumption for dielectric constant: $\varepsilon = \varepsilon'(f = 10 kHz)$. The significance of dielectric constant is related to the fact that it is directly responsible for the interaction of the given dielectric material with the external electric field. In dipolar materials, the test of $\varepsilon(T)$ temperature evolution shows the dominated spontaneous arrangement of permanent dipole moments, namely it is parallel for $d\varepsilon/dT < 0$ and antiparallel for $d\varepsilon/dT > 0$ [24].

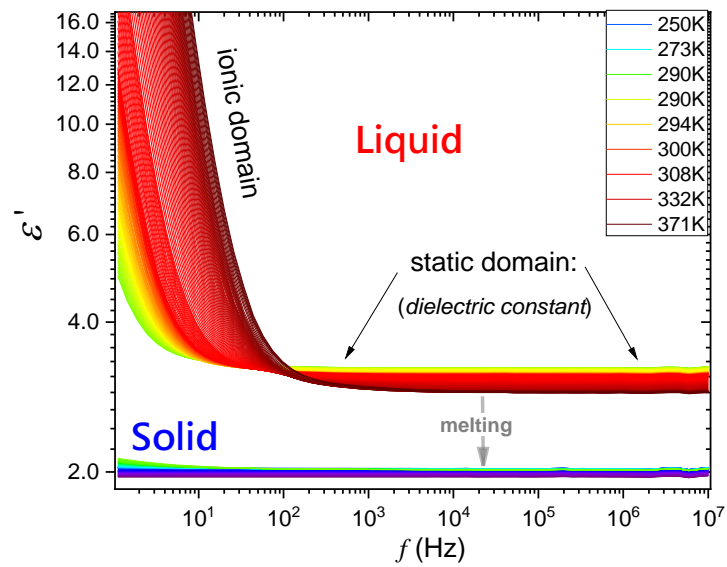

**Fig. 1** The frequency dependence of the of real part of dielectric permittivity in the tested temperature domain. Characteristic features of the spectrum are shown, including the manifestation of the melting phase transition.

Thymol has a small permanent dipole moment $\mu = 1.893D$ associated with O-H group in the molecular structure [25]. BDS studies were carried out using the capacitor from Invar, with the gap $d = 0.15mm$, and the voltage of the measurement electric field $U = 1V$.

The electrooptic Kerr effect (EKE) is the method associated with strong electric field refractive index birefringence, in respect to the direction of this field. The EKE setup was assembled in X-PressMatter Lab, and its scheme is shown in Fig. 2. The sample was placed in a cell containing two electrodes, distanced by 2 mm and 82 mm long. The cell's temperature was stabilized using Julabo thermostat with external circulation and V = 20 L of the cooling/heating agent liquid. Both in BDS and EKE studies, the temperature stabilization was better than 0.02K. Thymol was first heated to the liquid state and poured into the cell. Studies covered the domain below 355 K. On cooling, thymol remained liquid (supercooled liquid) when passing melting temperature and remained in supercooled liquid state until reaching the freezing temperature ~~282 K.

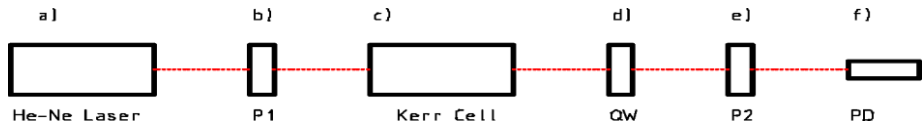

**Fig. 2**   Electrical Kerr Effect measurement setup: a) He-Ne Laser, b) polarizer P1, c) Kerr Cell, d) quarter-wave plate QW, e) polarizer P2, f) photo-diode PD

EKE facility was supported by Thorlabs' HRS015 Stabilized Red He-Ne Laser as a light source. Laser light traversed through polarizer P1 to ensure polarization with degree $α=π/4$ to the electric field. Next, the light beam traveling through Kerr cell (filled with sample material in-between capacitor plates), to which an electric field was applied. Following was a quarter-wave (QW) plate used as retarder [26] and polarizer P2 oriented at $α=-π/4$ (perpendicular to P1). Thorlabs' FDS010 photodiode (PD) was used as a light detector [24]. Photodiode current was amplified with Thorlabs' PDA200C Photodiode Amplifier and registered with Keysight DSOS104A oscilloscope. In tests, DC electric field pulses lasting typically 500 μs and voltage up to $U$ =1000 V were used. Their generation was supported by Agilent 33220A Function/Arbitrary Waveform Generator controlling DORA ZAS.2A 1000V power supply unit. For each measurement period photodiode current response was registered: before application of birefringence inducing electric field (for 1 ms), during application of electric field (for 0.5ms), and after removal of electric field (for 0.5 ms). At the output of EKE system, the laser beam signal was registered via photodiode, where current changes, reflected current sample properties under the electric field pulse, were registered. The response was normalized to the average value of current measured when no electric field was applied (before and after electric field removal). It is shown in Fig 3, together with the strong electric field pulse. Note the lack of the shift between baseliner before and after the pulse, which shows that no significant ionic contaminations that could yield a parasitic heating effect were present in tested samples. The normalized current response was then cumulated to improve the signal-to-noise ratio. The time-averaging (for further noise impact reduction) of cumulative current response (of photodiode was calculated.

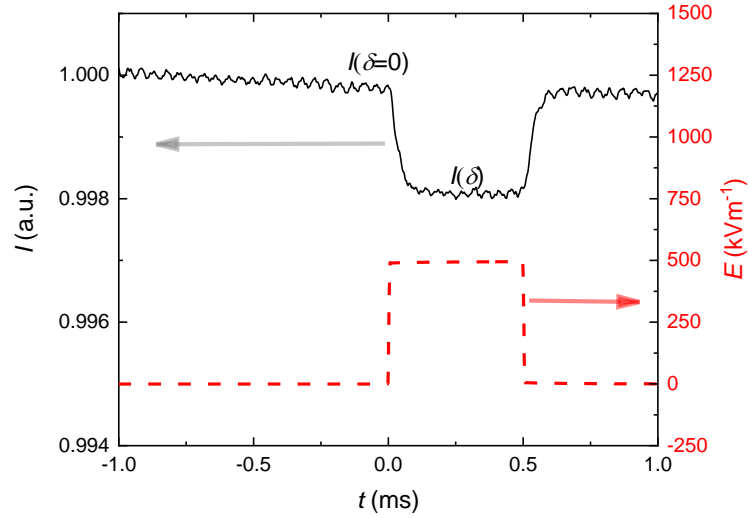

**Fig. 3** The shape of the detected photodiode current response (black, continuous line) of the sample to the pulse of the strong electric field (red, dashed line) measured at 327.3 K for impulses of 500 kV/m. Current values when no electric field was applied ($I(\delta=0)$) and when maximum electric field ($I(\delta)$) was applied are marked

The ratio of diode photocurrent while the field was not applied (and thus no birefringence was electrically induced $I(\delta=0)$) and when the field was applied (and photodiode signal was stable $I(\delta)$) was used to calculate phase shift between extraordinary and ordinary beams [27-31]:

$$\frac{\delta}{2} = arcsin\left(\sqrt{\frac{I(\delta)}{2 \cdot I(\delta = 0)}}\right) - \frac{\pi}{4} \tag{1}$$

where $\delta$ - phase shift between extraordinary and ordinary beams, $I(\delta=0)$ - current detected when no beam phase shift was electrically induced ('no' electric field baseline), $I(\delta)$ – electric current detected when maximal phase shift was observed (when the current response was stabilized).

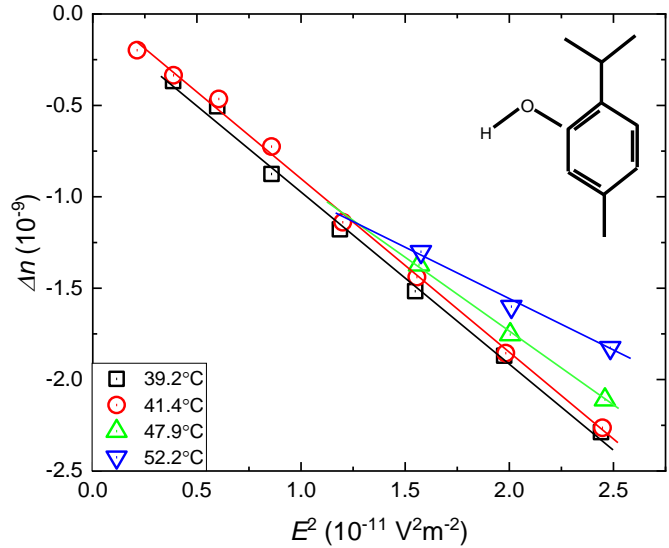

**Fig. 4** The plot shows the Kerr effect condition: anisotropy of the refractive index linear versus the square of electric field intensity in the tested sample at 312.3 K). On moving from the melting temperature, higher electric field were applied to obtain the response. The molecular structure of thymol is also shown

To support the registration, the setup contained the quarter-wave element, and the additional phase shift of $\pi/4$ was accounted in the estimation of the birefringence:

$$\Delta n = \delta \cdot \lambda / (\pi \cdot l) \quad \text{and} \quad B = \frac{\Delta n}{E^2} = \frac{\delta}{2 \cdot \pi \cdot l \cdot E^2} \tag{2}$$

where $\Delta n$ - birefringence, $\lambda$ - wavelength (in this case 633 nm), $\delta$ - phase shift between extraordinary and ordinary beams (eq. 1), $l$ - length of the optical path between capacitor plates (82 mm) in which birefringence was induced.

Birefringence is proportional to phase shift [32], which is induced with an electric field according to Kerr's law [33]. Linear dependence of EKE-related changes of the birefringence index *Δn*, which should be proportional to the square of the electric field intensity ('Kerr condition') [34]. The experimental validation of the 'Kerr condition' $\Delta n \propto E^2$ is shown in Figure 4. When comparing with other materials from the same homologous series (menthol), tested recently by the authors [14], one may observe the difference in electrically induced birefringence within the sign while values remain similar order of magnitude at low temperatures, with an increasing difference at higher temperatures (for the electric field of 500 kVm$^{-1}$ it is -2.29·10$^{-9}$ in thymol compared to 2.29·10$^{-9}$ in menthol at circa 312 K or -1.5·10$^{-9}$ in thymol and 0.96·10$^{-9}$ in menthol at circa 348 K).

3. **Results and Discussion**

Figure 5 shows the temperature evolution of dielectric constant, in the liquid and solid phases of thymol, on cooling and heating. The melting temperature is determined by heating from the solid phase $T_m = 323.0K$.

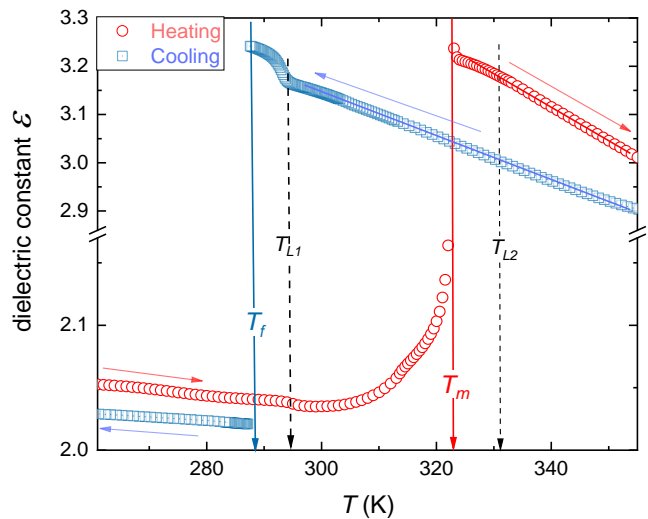

**Fig. 5**     The temperature dependence of dielectric constant on heating and cooling, as indicated by arrows. The melting and freezing discontinuous phase transition temperatures are indicated. Dashed vertical arrows indicate changes in the form of the temperature evolution.

On cooling from the liquid phase, supercooling down to $T_f = 288.0K$ occurred. Linear changes of dielectric constant, shown as solid lines, related to $d\varepsilon/dT < 0$, indicate that dipole moment follows the external electric field. It can be recognized as the hallmark of the 'parallel' arrangement in respect to the direction of the external electric field. However, the vicinity of $T_f$, the strong violation of the mentioned behavior occurs for $T_f < T < T_{L1} = T_f + 10K$. The distortion is detectable also on heating, although it is much weaker than for cooling. In the solid phase, the premelting effect emerges on heating from the solid phase. It is well manifested, but it is also associated with the high experimental resolution: the nonlinear rise from $\varepsilon(T_m - 15K) \approx 2.04$ to $\varepsilon \approx 2.5(T_m)$ takes place. Figure 6 shows the strong and long-range pretransitional changes of EKE in liquid thymol. It is presented in the scale testing the occurrence of the critical-like temperature evolution, earlier observed in refs. [12-14]:

$$B(T) = \frac{\Delta n}{\lambda E^2}(T) = \frac{A}{T-T^*} \quad \Rightarrow \quad B^{-1}(T) = A^{-1}(T - T^*) = aT - b \tag{3}$$

where $T < T_f$, $T_f$ denotes the freezing temperature, and $T^* < T_f$ is the singular, 'critical-like' temperature; constant parameters $a = A^{-1}T^*$, $b = A^{-1}$.

EKE changes detected on heating and cooling overlaps. The inset in Fig. 6 shows the distortions-sensitive and derivative-based validation of the behavior described by Eq. (3):

$$\frac{B^{-1}(T)}{dT} = A^{-1} \tag{4}$$

The inset also presents explicit evidence of the crossover ca. 10K above $T_f$, correlates with the temperature $T_{L1}$, detected in $\varepsilon(T)$ studies (Fig. 5). It is associated with different parameters in Eq. (3), namely: $A \approx -0.85 \, (10^{-14} mV^{-2})$, $T^* \approx 249K$, $\Delta T^*_{close} \approx 33K$, for $T_f < T < T_{L1}$, and $A \approx -9.2 \, (10^{-14} mV^{-2})$, $T^* \approx 188K$, $\Delta T^*_{far} \approx 94K$, for $T > T_{L1}$. The 'discontinuity is defined as $\Delta T^* = T_f - T^*$, The EKE pretransitional changes terminate at $\sim T_f + 45K$, which correlates with the value of $T_{L2}$ detected for dielectric constant (Fig. 5).

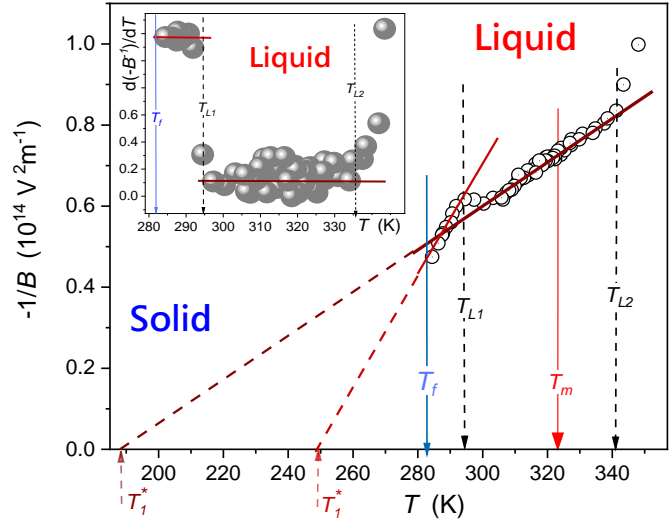

**Fig. 6** The temperature evolution of the negative reciprocal of the Kerr constant in liquid thymol. $B$ (birefringence) denotes the EKE constant. Indicated temperatures are associated with melting, freezing, and changes in temperature evolution. The parameterization is related to Eq. (3). The inset shows the derivative of experimental data from the central part of the plot, related to Eq. 4.

The unique feature of the pretransitional effect in Fig. 6 is the negative sign of the pretransitional effect, expressed by amplitudes $A < 0$, and opposite to the one noted in menthol. Notable that the 'positive' pretransitional anomaly, is the standard situation observed for instance, in the isotropic phase of LC materials, also described by Eq. (3), where it is associated with prenematic fluctuations, whose size and lifetime increase on cooling towards the Isotropic liquid (I) – Nematic (N) transition [10, 15-20]. It is the weakly discontinuous phase transition, for which the 'melting' of a single symmetry, uniaxial orientation, occurs. For pentylcyanobiphenyl (5CB), the classic rod-like LC material, pretransitional effects start at $T_{end} \approx T_{I-N} + 40K$ where fluctuations contain $2 - 3$ molecules. At $T \sim T_{\sim IN}$ this number rises to ca. ~ 100 molecules [15-20]. However, there is no mesophase between the liquid and crystalline states in menthol, and the model explanation based on the unique case of weakly discontinuous phase transitions in LC or PC materials [10, 12-20] cannot be applied.

For explaining the behavior described by Eq. (3) one can recall the model proposed by Hanus [35], predicting the creation of uniaxial filaments under the strong electric field in any molecular liquid near the melting temperature. Initially it was considered for $CS_2$ and nitrobenzene ($C_6H_5NO_2$), liquids without any tendency for creating mesophases. Hanus indicated that intermolecular coupling occurring in nitrobenzene essentially lowers the intensity of the electric field required for appearing filaments, in fact, to values comparable to ones applied in EKE studies. The first evidence for confirming Hanus predictions [35] experimentally was obtained in refs. [36, 37], for NDE studies in supercooled nitrobenzene. Hanus suggested that the phenomenon should be expected from $T_f$ to the domain slightly above $T_m$. Such characterization correlates with the behavior presented in Fig. 6. For Hanus model [35], molecules are specifically 'strung on an ordering thread' of a strong electric field, creating a filament. The amplitude in the Hanus model developed for EKE and NDE is proportional to the molecular anisotropy of the refractive index or dielectric constant, concerning the distinguished direction of the blue electric field. A strong electric field can be expected to organize the molecules based on the interaction with a dipole moment, and the intermolecular interaction associated with phenyl rings is a facilitating factor. For thymol molecules with slightly elongated structures and the dipole moment approximately perpendicular to the long axis, the birefringence describing changes of refractive index for the anisotropy induced by the electric field: $\Delta = n_\perp^E - n_{//}^E < 0$. The long molecular axis of thymol is perpendicular to the filaments axis, arranged by the direction of the strong electric fields. The 'positive sign' EKE anomaly ($A > 0$ in Eq. 3) in menthol suggests that the molecular axis characterized by the larger refractive index related polarizability is arranged along filaments in this material. The amplitude in Eq. 3: $A \propto (\Delta n \Delta \varepsilon)$ where the anisotropy $\Delta \varepsilon > 0$. Hence the sign of EKE pretransitional effect is governed by $\Delta n$.

Concluding, the report shows the existence of unique, 'negative-sign', pretransitional effects in liquid thymol. This result supplements the recent finding of the 'positive-sign' effect in menthol. The phenomenon can be explained in both cases via the Hanus model [35], suggesting the creation of molecular 'quasi-nematic' filaments structures under the strong electric field. Its existence is inherently associated with a strong electric field but can also reflect some generic features of the tested system. Worth stressing is the significant role of intermolecular interactions in Hanus model, which existence facilitates the creation of filaments. For thymol notable is the crossover detected in EKE and dielectric constant studies at $T_{L1}$. Strong changes in the evolution of dielectric constant in Fig. 5 indicate changes in the arrangement of permanent dipole moment, which can be associated with different local 'intermolecular' structures below and above $T_{L1}$. It associated with 'dramatic; changes of EKE evolution when passing $T_{L1}$, what shows the value of amplitude $A$ in Eq. (3). It allows posing the hypothesis for the liquid-liquid (L-L) transition [38-40] in liquid thymol at $T_{L1}$. L-L transition is a challenging phenomenon for which the direct experimental evidence is minimal [38-40 and refs therein]. The vast majority of results is related to indirect indications of L-L transitions hidden in hardly accessible experimental domains. Only recently, L-L was evidenced in a supercooled liquid just above the glass transition [40]. In the given report, the possible L-L transitions strongly manifest in only slightly supercooled thymol, in the easily accessible domain ~10 K above the freezing temperature ($T_f \approx 282\ K$).

This report supplements also recent communicates [12-14] regarding the occurrence of pretransitional effects for the still cognitively mystic melting/freezing discontinuous phase transition. From the practical point of view, the question also arises if the emergence of explicit critical-like behavior in liquid thymol, for a broad range of temperatures can open new possibilities for the liquid-based supercritical extraction technologies [41, 42] – based on the natural carrier, thymol?

**Acknowledgments**

Studies were carried out due to the National Centre for Science (NCN OPUS grant, Poland), ref.: 2017/25/B/ST3/02458, the head: S. J. Rzoska. The paper is associated with the *International Seminar on Soft Matter & Food – Physicochemical Models & Socio-Economic Parallels*, *1$^{st}$ Polish-Slovenian Edition*, Celestynów, Poland, 22–23 Nov., 2021; directors: Dr. hab. Aleksandra Drozd-Rzoska (Institute of High Pressure Physics PAS, Warsaw, Poland) and Prof. Samo Kralj (Univ. Maribor, Maribor, Slovenia).

**Authors Contributions statement**

A.Sz.-Sz. performed Kerr effect measurements. A.Sz.-Sz., S. J. R., and A. D. R discussed the results, analyzed data, and wrote the main part of the manuscript.

**Competing Interest**

There are no competing interests for the authors of this report.

**Data Availability Statement**

They are available from authors on request.

**References**

1. E. Boeker, R. van Grondelle, Environmental Physics: Sustainable Energy and Climate Change (John Wiley & Sons, Ltd, 2011). https://doi.org/10.1002/9781119974178
2. M.S. Rahman, Handbook of Food Preservation (CRC press, 2007).
3. P. Atkins. J. de Paula, Atkin's Physical Chemistry (Oxford Univ. Press., Oxford, 2001)
4. G. Arhenbout, Melt Crystallization Technology (Taylor & Francis, London, 2021).
5. W.C. Evans, D. Evans, Pharmacognosy (W.B. Saunders, London, 2009). https://doi.org/10.1016/B978-0-7020-2933-2.00022-8

6. Y. Yang, M. Asta, B.B. Laird, Solid-liquid interfacial premelting, Phys. Rev. Lett. **110** (2013) 096102. https://doi.org/10.1103/PhysRevLett.110.096102

7. A. Samanta, M. E. Tuckerman, Y, Tang-Qing, E, Weinan, Microscopic mechanisms of equilibrium melting of a solid, Sci. Rep. **346**, 729–732 (2014). https://doi.org/10.1126/science.1253810

8. V.P. Skripov, M.Z. Faizulin, Crystal-Liquid-Gas Phase Transitions and Thermodynamic Similarity (Wiley-VCH, Berlin, 2006).

9. Q.S. Mei, K. Lu, Melting and superheating of crystalline solids: from bulk to nanocrystals, Prog. Mat. Sci. **52** (8) (2007) 1175–1262.

10. M.A. Anisimov, Critical Phenomena in Liquids and Liquid Crystals (Gordon and Breach,m Reading, 1996).

11. J. Honig, J. Spalek, A Primer to The Theory of Critical Phenomena (Elsevier, Amsterdam, 2018).

12. A. Drozd-Rzoska, S. Starzonek, S.J. Rzoska. J. Łoś, Z Kutnyak, S. Kralj, Pretransitional effects of the isotropic liquid–plastic crystal transition, Molecules, **26**, 429 (2021).

13. A. Drozd-Rzoska, S. J. Rzoska, A. Szpakiewicz-Szatan, J. Łoś, S. Starzonek, Supercritical anomalies in liquid ODIC-forming cyclooctanol under the strong electric field, J. Mol. Liq. **345**, 1178491 (2022).

14. A. Drozd-Rzoska, S.J. Rzoska, A. Szpakiewicz-Szatan, J. Łoś, K. Orzechowski, Pretransitional and premelting effects in menthol, Chem. Phys. Lett. **793**, 139461 (2022).

15. A. Drozd-Rzoska, S.J. Rzoska, J. Ziolo, J. Jadżyn, Quasicritical behavior of the low-frequency dielectric permittivity in the isotropic phase of liquid crystalline materials, Phys, Rev. E **63**, 052701 (2001). https://doi.org/10.1103/PhysRevE.63.052701

16. A. Drozd-Rzoska, S.J. Rzoska, Complex relaxation in the isotropic phase of n-pentylcyanobiphenyl in linear and nonlinear dielectric studies, Phys. Rev. E **65**, 041701 (2002).

17. S.J. Rzoska, M. Paluch, S. Pawlus, A. Drozd-Rzoska, J. Ziolo, J. Jadzyn, k. Czupryńśki, R. Dąbrowski, Complex dielectric relaxation in supercooling and superpressing liquid-crystalline chiral isopentylcyanobiphenylM Phys. Rev. E **68**, 031705 (2003).

18. A. Drozd-Rzoska, S.J. Rzoska, S. Pawlus, J. Zioło, Complex dynamics of supercooling n-butylcyanobiphenyl (4CB), Phys. Rev. E. **72**, 031501 (2005). https://doi.org/10.1103/PhysRevE.72.031501

19. A. Drozd-Rzoska, S.J. Rzoska, M. Paluch, S. Pawlus, J. Zioło, P.G. Santangelo, C.M. Roland, Mode coupling behavior in glass-forming liquid crystalline isopentylcyanobiphenyl, Phys. Rev. E **71**, 011508 (2005).

20. S.J. Rzoska, A. Drozd-Rzoska, P. Mukherjee, D. Lopez, J. Martinez-Garcia, Distortion-sensitive insight into the pretransitional behavior of 4- n -octyloxy-4′-cyanobiphenyl (8OCB), J. Phys. Cond. Matt . **25**, 245105 (2013). https://doi.org/10.1088/0953-8984/25/24/245105

21. M.Y. Memar, P. Raei, N. Alizadeh, M. Akbari Aghdam, H.S. Kafil, Carvacrol and thymol: strong antimicrobial agents against resistant isolates, Rev. Med. Microbiol. **28**, 63-68 (2017). https://doi.org/10.1097/MRM.0000000000000100

22. H.J.D. Dorman, S.G. Deans, Antimicrobial agents from plants: antibacterial activity of plant volatile oils, J. Appl. Microbiol. **88**, 308–316 (2000). https://doi.org/10.1046/j.1365-2672.2000.00969.x

23. M. Trivedi, S. Patil, R. Mishra, S. Jana, Structural and physical properties of biofield treated thymol and menthol, Mol. Pharm. & Org. Proc. Res. **3**, 127 (2015). https://doi.org/10.4172/2329-9053.1000127

24. A. Chełkowski, Dielectric Physics (Elsevier-PWN, Warsaw, 1990).

25. R. Gupta A theoretical study of 5-methyl-2-isopropylphenol (thymol) by DFT, Int J. Res Sci & Technol. **8**, 812-830 (2021).

26. A. Mori, R. Tomita, Semi-automated sènarmont method for measurement of small retardation, Instrum. Sci. & Technol.**43**, 378-389 (2014). https://doi.org/10.1080/10739149.2014.1003072

27. E.D. Baily, B.R. Jennings, An apparatus for measurement of electrically induced birefringence, linear dichroism, and optical rotation of macromolecular solutions and suspensions, J. Coll. Inter.Inerfac. Sci. **45**, 177–189 (1973). https://doi.org/10.1016/0021-9797(73)90254-3

28. R. Piazza, V. Degiorgio, T. Bellini, Kerr effect in binary liquid mixtures, J. Opt. Soc. Am. B **3**, 1642-1646 (1986). https://doi.org/10.1364/JOSAB.3.001642

29. R.E. Hebner, R.J. Sojka, E.C. Cassidy, Kerr Coefficients of Nitrobenzene and Water, (U.S. Department of Commerce, Albuquerque, 1974).

30. W. Pyzuk, H. Majgier-Baranowska, J. Ziolo, Kerr effect in the critical solutions of Decyl Alcohol, Chem. Phys. **59**, 111–118 (1981).

31. W. Pyzuk, I. Słomka, Determination of the Kerr constant in optically active liquids and chiral discrimination in menthone, J. Phys. D: Appl. Phys. **17**, 171 (1984). https://doi.org/10.1088/0022-3727/17/1/023

32. D.P. Shelton, High accuracy Kerr effect measurement technique, Rev. Sci. Instrum. 64, 917–931 (1993). https://doi.org/10.1063/1.1144144

33. H. Watanabe, A. Morita, Kerr effect relaxation in high electric fields, Adv. Chem. Phys. **56**, 255–409 (1984).

34. A.W. Knudsen, The Kerr Effect in nitrobenzene–a student experiment, Am. J. Phys. **43** (1975) 888–894.

35. J. Hanus, Effect of the molecular interaction between anisotropic molecules on the optical kerr effect. field-induced phase transition, Phys. Rev. **178**, 420–428 (1969).

36. A. Drozd-Rzoska, S. J. Rzoska, J. Zioło, Anomalous temperature behavior of nonlinear dielectric effect in supercooled nitrobenzene, Phys. Rev. E **77**, 041501 (2008).

37. A. Drozd-Rzoska, S.J. Rzoska, A.A. Rzoska, Pretransitional behavior of the nonlinear dielectric effect for the liquid-solid transition in nitrobenzene, Phys. Rev. E. **93** (2016) 062131. https://doi.org/10.1103/PhysRevE.93.062131

38. M. Kobayashi, H. Tanaka, H. The reversibility and first-order nature of liquid-liquid transition in a molecular liquid, Nat. Comm. **7**, 1–8 (2016).

39. M. Mierzwa, M. Paluch, S.J. Rzoska, J. Zioło, the liquid− glass and liquid− liquid transitions of TPP at elevated pressure, J. Phys. Chem. B **112**, 10383-10385 (2008).

40. Z. Wojnarowska, S. Cheng, B. Yao, M. Swadzba-Kwasny, S. McLaughlin, A. McGrogan, Y. Delavoux, M. Paluch, Pressure-induced liquid-liquid transition in a family of ionic materials, Nat. Comm. **13**, 1342 (2022).

41. T. Fornari, R.P. Stateva, High Pressure Fluid Technology for Green Food Processing (Springer, Heidelberg, 2015).

42. S.J. Rzoska and A. Drozd-Rzoska, Criticality-related fundamental bases for new generations of gas-liquid, liquid-liquid, and liquid (LE) extraction technologies, Available at: https://arxiv.org/ftp/arxiv/papers/2207/2207.01668.pdf